\documentstyle[twoside,fleqn,espcrc2]{article}

\newcommand{\AmS}{{\protect\the\textfont2
  A\kern-.1667em\lower.5ex\hbox{M}\kern-.125emS}}

\title{Topological properties of full QCD at the phase transition}

\author{B. All\'es\thanks{Speaker at the conference.}\address{Dipartimento 
                           di Fisica, Sezione Teorica,
                           Universit\`a di Milano, Via Celoria 16,
                           20133 Milano, Italy},
        M. D'Elia$^{\rm b}$, A. Di Giacomo$^{\rm b}$, 
        P. W. Stephenson\address{Dipartimento
                  di Fisica, Universit\`a di Pisa, Piazza Torricelli 2,
                  56126 Pisa, Italy}}
\begin{document}

\begin{abstract}
We investigate the topological
properties of the QCD vacuum with 4 flavours of
dynamical staggered fermions at finite temperature. 
To calculate the topological susceptibility we use the field--theoretical
method. As in the quenched case, a sharp drop
is observed for the topological susceptibility across the phase transition.
\end{abstract}

\maketitle

\section{INTRODUCTION}

The topological susceptibility is defined as 
\begin{equation}
 \chi \equiv \int d^4x \partial_\mu \partial_\nu 
   \langle {\rm T}\left(K_\mu(x) K_\nu(0)\right) \rangle
\label{eq:chi}
\end{equation}
where $K_\mu(x)$ is 
\begin{eqnarray}
 K_\mu(x) &=& {g^2 \over 16 \pi^2} \epsilon_{\mu\nu\rho\sigma} \times \nonumber \\
  & &A_\nu^a \left( \partial_\rho A^a_\sigma - {1 \over 3} 
  g f^{abc} A_\rho^b A_\sigma^c \right)
\end{eqnarray}
and $Q(x)=\partial_\mu K_\mu(x)$ is
the density of topological charge.
Eq.~(\ref{eq:chi}) uniquely defines the prescription for the singularity 
of the time ordered product at $x \longrightarrow 0$~\cite{witten}.
Determining $\chi$ around $T_c$ in full QCD is an important piece of
information to test models of QCD vacuum~\cite{shuryak}.

\subsection{The Simulation}

\begin{figure}[htb]
\vspace{4.5cm}
\includegraphics{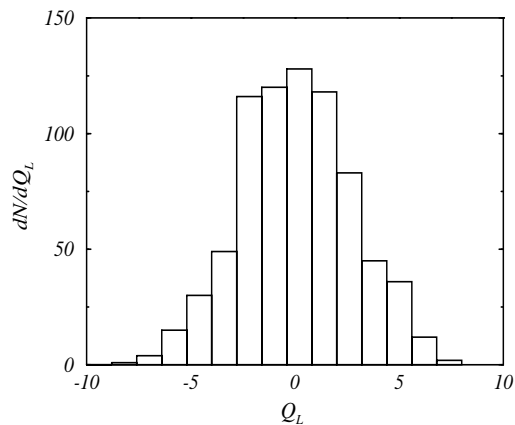}
\null\vskip 0.3cm
\caption{Distribution of topological charge $Q_L$ at $\beta=5.04$.}
\label{fig:qdecorrel}
\end{figure}

We have simulated the theory on a $16^3\times 4$ lattice with four flavours
of staggered fermions with bare mass $am=0.05$. We have used the HMC 
algorithm for the updating. 
To be sure that topology is well thermalized we have checked the 
topological charge of our sample of configurations by cooling.
They are well decorrelated.
In Fig.~\ref{fig:qdecorrel} we display the 
distribution of topological
charge of our configurations at 
$\beta\equiv 6/g^2=5.04$. This 
histogram confirms that at this mass value 
the topology is well sampled~\cite{boyd}.
We have learned that topology is easily decorrelated when we
work with large quark masses~\cite{boyd,lippert}. In our case 
also the lattice spacing is large, so that updating the instanton
content is faster. 

\begin{figure}[htb]
\vspace{4.5cm}
\includegraphics{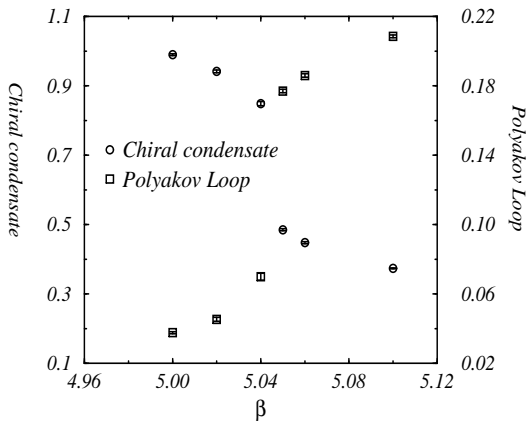}
\null\vskip 0.3cm
\caption{Polyakov loop (squares, right scale) and chiral condensate 
(circles, left scale) across the phase transition
on a $16^3\times 4$ lattice.}
\label{fig:polyakov}
\end{figure}

With our bare fermion mass and lattice size the deconfining 
transition is known to occur at 
$\beta_c = 5.04$~\cite{brown}. We have checked this number
by computing the Polyakov loop and the chiral condensate as shown in 
Fig.~\ref{fig:polyakov}.


The topological susceptibility was measured 
at $\beta$=5.00, 5.02, 5.04, 5.05, 5.06, 5.10.

\begin{table*}[hbt]
\setlength{\tabcolsep}{5.0pc}
\newlength{\digitwidth} \settowidth{\digitwidth}{\rm 0}
\catcode`?=\active \def?{\kern\digitwidth}
\caption{Topological susceptibility and $T/T_c$ vs. $\beta$.}
\label{tab:tab}
\begin{tabular*}{\textwidth}{@{}l@{\extracolsep{\fill}}rrrr}
\hline
&                   \multicolumn{1}{r}{$\beta$} 
                 & \multicolumn{1}{r}{$T/T_c$} 
                 & \multicolumn{1}{r}{$\chi$/MeV$^4$}         \\
\hline
&5.00 & 0.9677 & 1.61(43) \\
&5.02 & 0.9804 & 1.13(30) \\
&5.04 & 1.0000 & 1.21(30) \\
&5.05 & 1.0126 & 2.89(1.46) \\
&5.06 & 1.0274 & 1.63(1.22) \\
&5.10 & 1.1110 & 0.43(1.19) \\
\hline
\end{tabular*}
\end{table*}

\subsection{The Operators}

We have measured the topological susceptibility by using 
the 2--smeared lattice topological charge $Q_L$~\cite{haris}
\begin{equation}
 Q_L(x) = -{1 \over 2^9 \pi^2} \sum \epsilon_{\mu\nu\rho\sigma}
          {\rm Tr}\left( \Pi_{\mu\nu}(x) \Pi_{\rho\sigma}(x) \right)
\label{eq:ql}
\end{equation}
and the field--theoretical method~\cite{tanti} to extract $\chi$ from
the lattice susceptibility
\begin{equation}
 \chi_L \equiv {Q_L^2 \over V} = Z^2 a^4 \chi + M
\label{eq:chil}
\end{equation}
where $Q_L$ is the lattice total charge $Q_L\equiv \sum_x Q_L(x)$ and $V$ the 
space--time volume. The multiplicative and additive 
renormalizations $Z$ and $M$ have been 
evaluated by use of the heating method~\cite{tanti2,tsu3}.

\section{DETERMINATION OF $a$, $Z$ AND $M$}

The lattice spacing was extracted by measuring the string tension 
at $\beta=5.00$, 5.04 and 5.10 on 
a $16^4$ lattice. For the other values of $\beta$ it was determined
by a splines interpolation.
Wilson loops were evaluated by using smeared spatial links. 
From the value at $\beta_c$ ($a = 0.300(20) {\rm fm}$) we infer the critical 
temperature to be $T_c=164(11)$ MeV. 

The heating method to evaluate $Z$ and $M$ yields a non--perturbative 
determination of these renormalization constants~\cite{tanti2,tsu3}.
To calculate $Z$ a local updating algorithm 
is applied on a configuration containing 
a charge +1 classical instanton. These updatings, being local, thermalize
the short distance fluctuations, responsible for the renormalization effects,
and due to the slowing down leave large structures, like instantons, unchanged.
The measurement of $Q_L$ on such updated configurations yields $Z \cdot Q$. As
$Q$ is known, one can extract $Z$. Notice that this procedure is equivalent
to imposing the continuum value for the 1--instanton charge (in the
$\overline{\rm MS}$ scheme it is +1) and extracting the finite multiplicative
renormalization $Z$ by evaluating a matrix element of $Q_L$.

The additive renormalization $M$ is obtained in a similar way. We
apply a few heating steps with a local updating algorithm on a zero--field
configuration. Then the topological susceptibility is calculated.
This provides the value of $M$, if no instantons have been created during
the few updating steps. This
method agrees with the treatment of the singularity of~Eq.(\ref{eq:chi}).

As explained in~\cite{tsu3}, cooling tests must be done to check that the
background topological charge has not been changed during the local heating
(it is +1 in the calculation of $Z$ and 0 in the evaluation of $M$).

\section{RESULTS}

\begin{figure}[htb]
\vspace{4.5cm}
\includegraphics{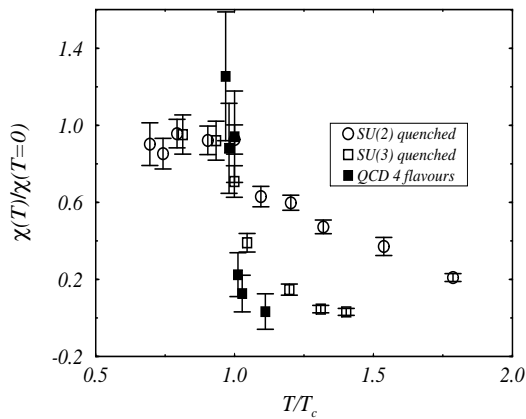}
\null\vskip 0.3cm
\caption{Behaviour of the topological susceptibility as a function
of the normalized temperature $T/T_c$.}
\label{fig:chiT}
\end{figure}

The values for the topological susceptibility and the 
temperature are given in table~\ref{tab:tab}.
In Fig.~\ref{fig:chiT} we show the normalized topological susceptibility 
$\chi(T)/\chi(T=0)$ as a function of 
the temperature. The value at zero temperature $\chi(T=0)$ has been
obtained by averaging the values at $T < T_c$.
In the same figure the results for the 
quenched case~\cite{tsu3,tsu2} are shown for comparison.
The signal for $\chi$ drops strongly when crossing the 
transition temperature. Analogous results have been reported for two flavours
and by using the cooling method~\cite{deforcrand}. 

From our data we cannot 
yet conclude that the drop in presence of fermions 
is steeper than for the 
quenched case, especially after considering that 
our simulations have been done at low beta values, where
a poor scaling is expected, with a possible resulting
large systematic effect on the scale $T/T_c$. 

Comparing the full QCD results to the quen-\break
ched case is an important
issue for testing instanton liquid models of the vacuum. In view
of that we are adding determinations
on a $32^3\times8$ lattice where we expect better control on scaling.

\section{Acknowledgements}

A. Di Giacomo acknowledges the financial contribution of the 
European Commission under the TMR-Program ERBFMRX-CT97-0122.

\end{document}